\documentclass[a4paper,11pt]{article}
\usepackage{pos}
\usepackage{amsmath}
\usepackage{bm}

\title{Probing the space--time structure of the jet-induced medium response with HBT interferometry}
\ShortTitle{HBT interferometry of the jet-induced medium response}

\author*[a]{Dexing Zhu}
\author[b]{Jinhui Chen}
\author[a]{Defu Hou}
\author[a]{Weiyao Ke}
\author[a]{Jiahao Shi}
\author[b]{Baoshan Xi}
\author[c]{Zhong Yang}
\author[b]{Chunjian Zhang}

\affiliation[a]{Key Laboratory of Quark and Lepton Physics (MOE) \& Institute of Particle Physics, Central China Normal University, Wuhan 430079, China}

\affiliation[b]{Key Laboratory of Nuclear Physics and Ion-beam Application (MOE) \& Institute of Modern Physics, Fudan University, Shanghai 200433, China}

\affiliation[c]{Department of Physics and Astronomy, Vanderbilt University, Nashville, TN 37235, USA}

\emailAdd{weiyaoke@ccnu.edu.cn}
\emailAdd{houdf@ccnu.edu.cn}

\abstract{Energetic partons traversing the quark--gluon plasma (QGP) deposit energy and momentum in the medium and excite a medium response. 
The emergence of Mach-cone and diffusion-wake structures is a signature that this response arises from collective dynamics. 
However, these space-time structures as a result of jet perturbation is strongly washed out during the freeze out process.
To probe the QGP response behavior in real space, we propose to study the Hanbury Brown--Twiss (HBT) interferometry of identical particles in events triggered by a hard process.
We use the \textsc{CoLBT-hydro} model to generate $\gamma$-triggered jet events in $0$--$10\%$ O+O collisions at $\sqrt{s_{\rm NN}}=5.36$~TeV, and apply the Lednick\'y--Lyuboshits formalism to compute HBT correlation function. 
First, we compare $\pi^+\pi^+$ correlation functions with and without the jet-induced medium response in the simulation, excluding particles associated with MPI and jet fragmentation.
The response alters the correlation function in the $\gamma$ side by about $0.2\%$ at relative momenta $k^\ast\approx0.03$--$0.08$~GeV/$c$.
Nevertheless, further including particle production from MPI and jet fragmentation produces a much larger modification at $k^\ast\lesssim0.06$~GeV/$c$, which must be understood with better accuracy and develop corresponding subtraction techniques to isolate the residue collective response signal.
}

\FullConference{
13th International Conference on Hard and Electromagnetic Probes of High-Energy Nuclear Collisions\\
(Hard Probes 2026)\\
21-26 June 2026\\
Nashville, Tennessee, USA
}

\begin{document}
\maketitle

\section{Introduction}
A fast parton traversing the quark--gluon plasma (QGP) deposits energy and momentum in the medium and is expected to induce a collective response due to the fuild behavior of QGP\cite{Qin2009,Neufeld2008,Chen2018,Yang2021,Yang2023}.
The perturbed medium exhibits two characteristic features: a Mach cone, namely a conical sound front, along the direction of jet propagation, and a diffusion wake, corresponding to a depletion of the medium behind the jet.
Both of these phenomena are clearly seen in the space-time profile of the QGP, but experimentally one can only observe particle correlations in the momentum space. 

In momentum space, simulations have demonstrated that collective response modifies the soft hadrons yield correlated with the jet, leading, e.g., to enhancement of jet shape at large angles~\cite{Tachibana2017} and $\gamma$/$Z$-hadron correlations at small transverse momentum~\cite{Chen2018,Yang2021} relative to the baselines in proton-proton collisions.
Even so, it can be hard to address the nature of such response if we are solely based on the observation in momentum space because both particle-like excitation and collective excitation can end up with soft-hadron enhancement, while they have completely different response functions in space and time.
The reason behind this is that the freeze-out of the momentum space largely washes out the detailed correlations between the medium response structure and the momentum space observables.

To probe residual information related to the space-time structure of the collective response, we propose to measure the Hanbury Brown--Twiss (HBT) interferometry of identical particles~\cite{HBT1956,GGLP1960} in a sample of events triggered by jets.
HBT has been used to measure the extent and inhomogeneity of the space-time of the emitting source~\cite{Lisa2005,Plumberg2018}. 
 The underlying idea is that if a jet perturbs the medium and creates a characteristic conical collective response, it should modify the event-by-event inhomogeneity of the emitting source and thereby leave an imprint on the HBT correlation function. 
 This approach therefore offers a complementary, coordinate-space probe of the jet-induced response.

In this contribution, we use the \textsc{CoLBT-hydro} model~\cite{He2015,Pang2012,Pang2018} to generate $\gamma$-triggered jet events in central O+O collisions and apply the Lednick'y--Lyuboshits (LL) formalism~\cite{LL1981,Lednicky2009,LL2000,Xi2025} to calculate the $\pi^+\pi^+$ HBT correlation function. 
As a preliminary study, we first test the impact of including only particles produced from the jet-modified freeze-out hypersurface and then,  including particle from jet fragmentation and MPI associated to the hard processes.

\section{Methods}
For a chaotic source and in the absence of final-state interactions, the correlation function of two identical bosons with four moments $p_1$ and $p_2$ reads~\cite{Lisa2005}
\begin{equation}
  C(q,K)=\frac{E_1E_2\,\mathrm{d}N/\mathrm{d}^3p_1\mathrm{d}^3p_2}{(E_1\,\mathrm{d}N/\mathrm{d}^3p_1)(E_2\,\mathrm{d}N/\mathrm{d}^3p_2)}
  \simeq 1+\frac{|\tilde S(q,K)|^2}{|\tilde S(0,K)|^2},
  \qquad
  \tilde S(q,K)\equiv\int\mathrm{d}^4x\,S(x,K)\,e^{iq\cdot x},
  \label{eq:C}
\end{equation}
where $q=p_1-p_2$, $K=(p_1+p_2)/2$, and $S(x,K)$ is the emission function of the source.
In the pair rest frame $q=(0,2\bm k^\ast)$, so the correlation function at relative momentum $k^\ast$ probes source structures of size $\sim\hbar /(2k^\ast)$, up to a Lorentz boost from the pair rest frame back to the laboratory frame.
For an event with a jet, we decompose the emission function as
\begin{equation}
  S(x,K)=S_{\rm bg}(x,K)+S_{\rm resp}(x,K)+S_{\rm jet}(x,K),
  \label{eq:decomp}
\end{equation}
where $S_{\rm bg}$ describes the unperturbed medium, $S_{\rm resp}$ its modification by the jet-induced response, and $S_{\rm jet}$ the emission function from the hard process, including jet fragmentation, beam remnant, and the associated MPI activity.
Expanding $|\tilde S|^2$, we organize the terms into diagonal contributions and interference
\begin{align}
|\tilde{S}|^2 &= |\tilde S_{\rm jet}|^2 + |\tilde S_{\rm bg}|^2 + |\tilde S_{\rm resp}|^2 \nonumber\\
&~ +  2\,\mathrm{Re}\big(\tilde S^\ast_{\rm bg}\tilde S_{\rm resp}\big)
  +2\,\mathrm{Re}\big(\tilde S^\ast_{\rm bg}\tilde S_{\rm jet}\big)
  +2\,\mathrm{Re}\big(\tilde S^\ast_{\rm resp}\tilde S_{\rm jet}\big)\nonumber\\
&\approx |\tilde S_{\rm jet}|^2 + |\tilde S_{\rm bg}|^2  + 2\,\mathrm{Re}\big(\tilde S^\ast_{\rm bg}\tilde S_{\rm resp}\big)
  +2\,\mathrm{Re}\Big(\big(\tilde S^\ast_{\rm bg}+\tilde S^\ast_{\rm resp}\big)\tilde S_{\rm jet}\Big)\,.
  \label{eq:cross}
\end{align}
Since $\tilde{S}_{\rm resp}$ is small compared with $\tilde{S}_{\rm bg}$, we have dropped the $|\tilde{S}_{\rm resp}|^2$ in the last line.

To obtain $S(x,K)$, we simulate $0$--$10\%$ O+O collisions at $\sqrt{s_{\rm NN}}=5.36$~TeV with and without a photon trigger of $p_T^\gamma>100$~GeV/$c$. 
To incorporate quantum interference effects into the correlation function, we employ the Lednick\'y--Lyuboshits (LL) formalism~\cite{Lednicky2009,LL2000,Xi2025}, which also accounts for Coulomb and strong final-state interactions between the two particles.
The correlation functions are constructed from same-event and mixed-event pair distributions, as in experimental analyses (``exp.\ method'' in the figures), and are shown as functions of the relative momentum $k^\ast$ in the pair rest frame.
``Signal'' and ``background'' denote events with and without the jet-induced medium response, respectively.
The pions used for the analysis are selected within $|\eta-\eta_{\rm jet}|\leq2.0$ and $|\phi-\phi_\gamma|\leq1.0$.

\section{Results}
\label{sec:results}

\begin{figure}[htbp]
  \centering
  \includegraphics[width=.46\textwidth]{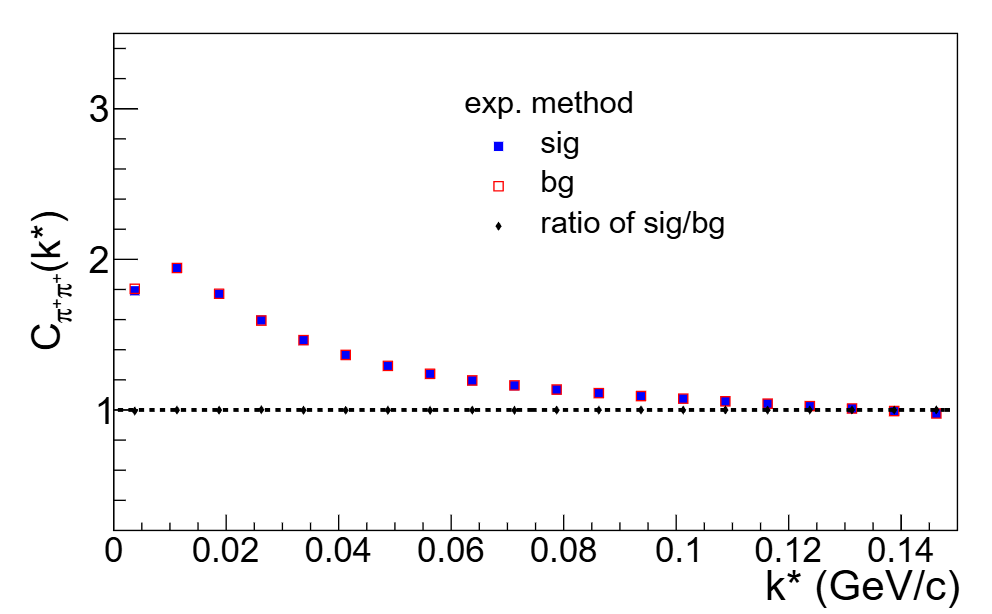}
  \hfill
  \includegraphics[width=.46\textwidth]{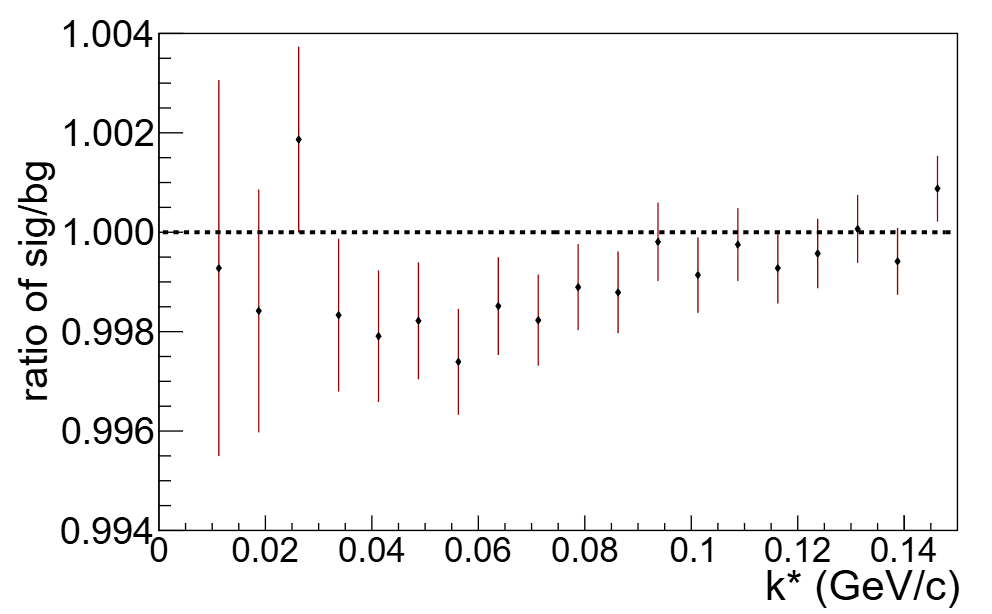}
  \caption{$\pi^+\pi^+$ correlation functions from the jet-modified freeze-out surface in $0$--$10\%$ central O+O collisions at $\sqrt{s_{\rm NN}}=5.36$~TeV with a $\gamma$-jet trigger ($p_T^\gamma>100$~GeV/$c$), calculated with the LL formalism. 
  Only pions emitted from the freeze-out surface are included; particles from jet fragmentation and MPI are excluded, so that the calculation isolates the effect of the jet-induced medium response. 
  Left: signal (with medium response) and background (without medium response) correlation functions, together with their ratio (black points). 
  Right: the same signal-to-background ratio shown on an expanded vertical scale around unity. 
  }

  \label{fig:ll_surface}
\end{figure}

\paragraph{Event-averaged result from the jet-modified freeze-out surface.}
To better understand the structure of the modification, we first investigate the difference between $|\tilde S_{\rm bg}|^2$ and $|\tilde S_{\rm bg}|^2  + 2\,\mathrm{Re}\big(\tilde S^\ast_{\rm bg}\tilde S_{\rm resp}\big)$, i.e., the difference caused by the interference between medium response and the unperturbed medium, while all the jet-related correlations are omitted. 
This is the contribution of interest because it directly probes the modification induced by the collective response. 
Figure~\ref{fig:ll_surface} shows the result when only the jet-modified freeze-out surface is included.
Their ratio reveals a systematic suppression of about $0.2\%$ for $0.03\lesssim k^\ast\lesssim0.08$~GeV/$c$, with a minimum near $k^\ast\simeq0.06$~GeV/$c$.
Such a typical $k^\ast$  corresponds to a spatial scale $r_0\sim\hbar c/(2k^\ast)\approx1.7$~fm in the rest frame of the pair. It should be interpreted as the typical change of inhomogeneity of the emitting source by medium response rather than the actual size of the Mach cone or the diffusion wake.

\paragraph{Including pions from jet fragmentation (preliminary result).}
We next add particles from the fragmentation of the jet partons evolved by LBT to evalate the sum of all terms in Eq.~\eqref{eq:cross}. It should be noted that in the LBT treatment, such jet particles also include those from the associated MPI process, while a more consistent treatment should include those mini jets as a change of the initial condition of the bulk evolution. Therefore, these calculation is a preliminary study.

The modification is now much larger due to the inclusion of jet fragmentation hadrons. The ratio decreases from unity at $k^\ast\approx0.06$~GeV/$c$ to about $0.9$ at $k^\ast\approx0.01$--$0.02$~GeV/$c$ and to about $0.4$ in the lowest bin, shows a small (${\approx}4\%$) enhancement near $k^\ast\approx0.07$~GeV/$c$, and approaches unity for $k^\ast\gtrsim0.1$~GeV/$c$.
The jet-associated pions thus dominate the low-$k^\ast$ signal over the response-induced correction of Fig.~\ref{fig:ll_surface}.
It is yet to be tested whether such modification will be altered if one properly treat the particles coming out of MPI as part of bulk evolution initial condition. But this points out that it is important to understand and subtract correlations associated to hardons from jet fragmentation to isolate the contribution from collective response for an extraction of the latter and inferring the nature of medium response.

\begin{figure}[htbp]
  \centering
  \includegraphics[width=.46\textwidth]{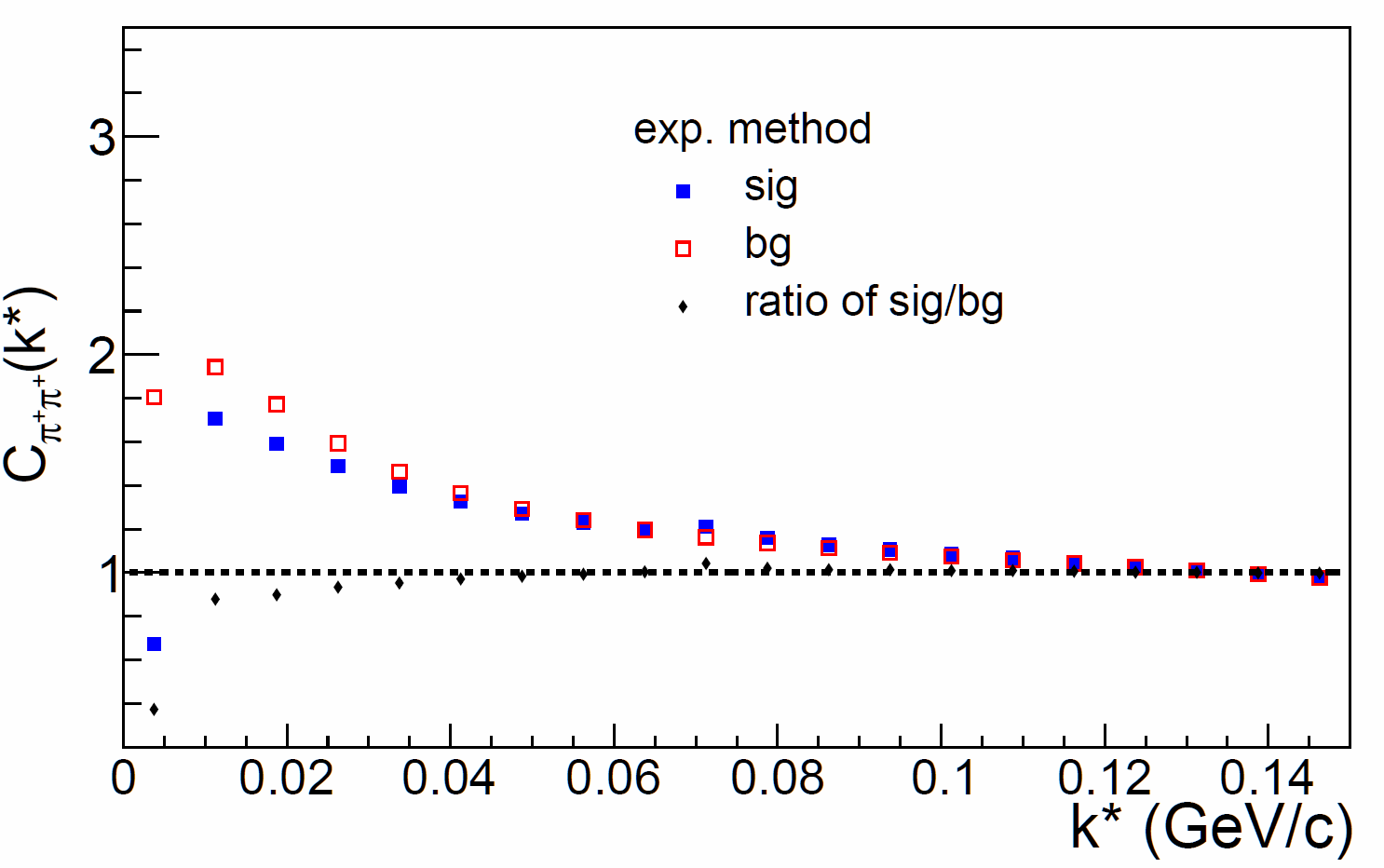}
  \hfill
  \includegraphics[width=.46\textwidth]{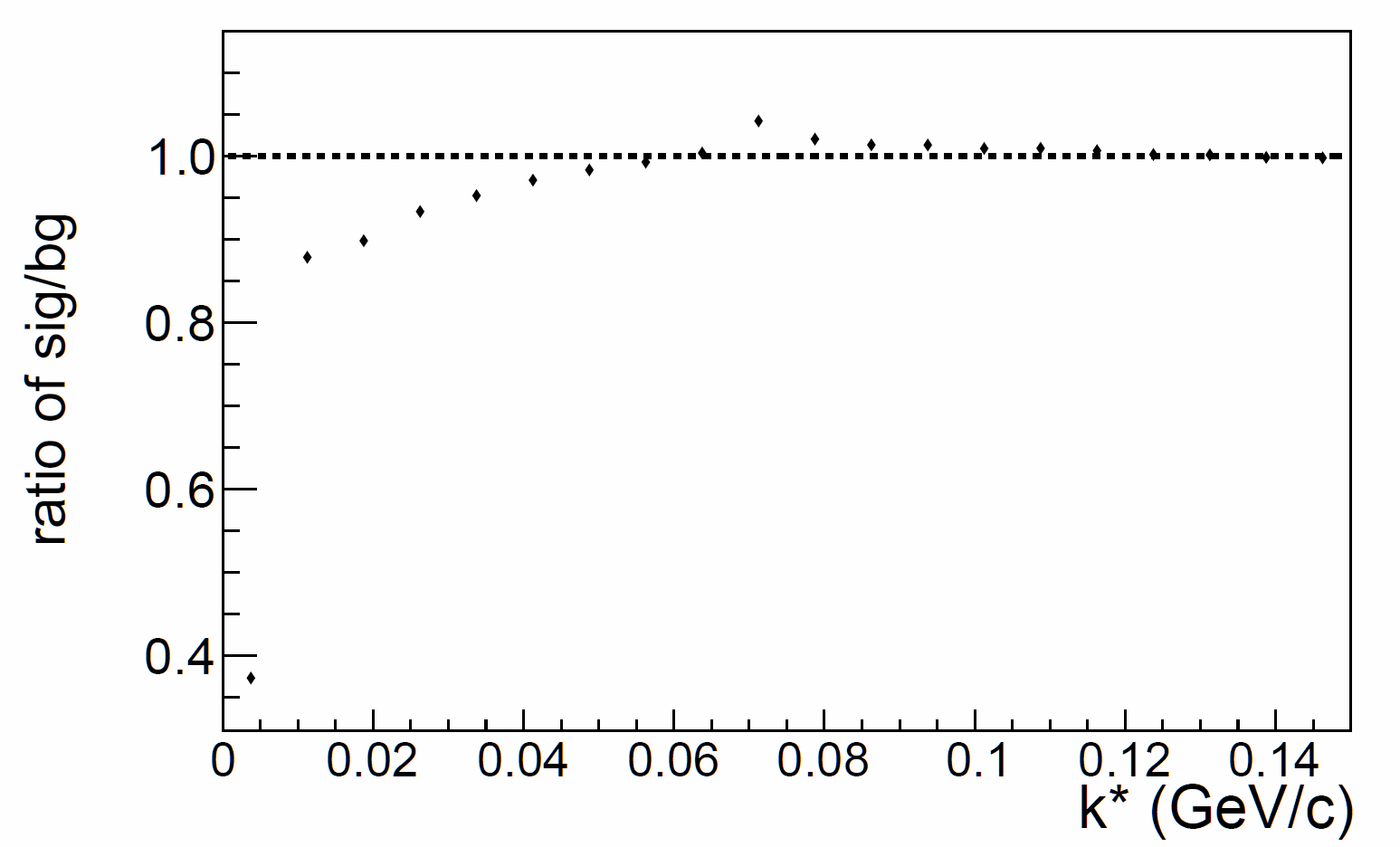}
  \caption{Same as Fig.~\ref{fig:ll_surface}, but with pions from jet fragmentation and associated MPI activity added to the signal (preliminary); the background is unchanged.
  Left: signal and background correlation functions, together with their ratio.
  Right: the signal-to-background ratio shown separately.}
  \label{fig:ll_jetmedium}
\end{figure}

\section{Summary}
\label{sum}

We have proposed HBT interferometry as a probe of the jet-induced medium response that retain a certain level of coordinate-space correlation.
For $\gamma$-triggered jets in O+O collisions at $\sqrt{s_{\rm NN}}=5.36$~TeV, the jet-modified freeze-out surface suppresses the $\pi^+\pi^+$ correlation function by about $0.2\%$ at $k^\ast\approx0.03$--$0.08$~GeV/$c$.
Pions from jet fragmentation produce a much larger modification at low $k^\ast$, which constitutes the major contamination to isolate the collective medium response signal.
A controlled subtraction of this contribution will be the key to extracting the former signal and to study the space--time structure of the medium response from final-state correlations.

\end{document}